\documentclass[secnumarabic,twocolumn,aps,prd,amsmath,amssymb,nofootinbib]{revtex4}
\bibpunct{}{}{,}{s}{}{,}
\usepackage{bm}

\usepackage{amsmath}
\usepackage{amsfonts}
\usepackage{amssymb}
\usepackage{graphicx}

\def\beq{\begin{equation}}
\def\eeq{\end{equation}}
\def\bea{\begin{eqnarray}}
\def\eea{\end{eqnarray}}

\newcommand{\kpc}{\,\mbox{kpc}}
\newcommand{\rtwo}{r_{200}}
\newcommand{\mtwo}{M_{200}}
\newcommand{\msun}{M_\odot}

\newcommand{\vmax}{V_{\rm max}}

\newcommand{\rvmax}{r_{\rm Vmax}}

\newcommand{\kms}{\,{\rm km\,s^{-1}}}
\newcommand{\sden}{\,{\rm M_\odot\,pc^{-3}}}
\newcommand{\psden}{\,{\rm M_\odot\,pc^{-3}\,km^{-3}\,s^{3}}}

\long\def\symbolfootnote[#1]#2{\begingroup%
\def\thefootnote{\fnsymbol{footnote}}\footnote[#1]{#2}\endgroup}

\begin{document}

\title{The structure and evolution of cold dark matter halos}

\author{J\"urg Diemand}
\email{diemand@ucolick.org}
\affiliation{Hubble Fellow}
\affiliation{Astronomy \& Astrophysics, University of California, Santa Cruz, CA 95060, U.S.A.}

\author{Ben Moore}
\email{moore@physik.uzh.ch}
\affiliation{Institute for Theoretical Physics, University of Z\"urich, CH-8057 Z\"urich, Switzerland}

\begin{abstract}
In the standard cosmological model a mysterious cold dark matter (CDM)
component dominates the formation of structures. Numerical studies of the formation of CDM halos have 
produced several robust results that allow unique tests of the hierarchical
clustering paradigm. Universal properties of halos, including their mass
profiles and substructure properties are roughly consistent with observational
data from the scales of dwarf galaxies to galaxy clusters. Resolving the fine
grained structure of halos has enabled us to make predictions for
ongoing and planned direct and indirect dark matter detection experiments.

While simulations of pure CDM halos are now very accurate and in good agreement
(recently claimed discrepancies are addressed in detail in this review), we are still unable
to make robust, quantitative predictions about galaxy formation and
about how the dark matter distribution changes in the process.
Whilst discrepancies between observations and simulations have been the subject of much debate in the literature,
galaxy formation and evolution needs to be understood in more detail in order to fully test the CDM paradigm.
Whatever the true nature of the dark matter particle is, its clustering properties must not be too different from 
a cold neutralino like particle to maintain all the successes of the model in matching large scale structure data and the global
properties of halos which are mostly in good agreement with observations.

%Within these uncertainties the structure of CDM halo seems consistent
%with observations. There are some apparent tensions on small scales, mainly: 
%i) the huge abundance of CDM subhalos vs. few dwarf satellite galaxies
%ii) cuspy DM density profiles vs.  slowly rising galaxy rotation curves. But galaxies
%need to be understood in detail before one might rule out CDM based on such
%indirect probes. On the other hand, some non-gravitational evidence
%(i.e. detecting DM particles or their decay/annihilation products)
%would be required to prove the existence of dark matter.

\bigskip

\end{abstract}

\maketitle

\tableofcontents

\newpage

\section{Introduction: From cold collapse to hierarchical clustering}

\subsection{A short history}

N-body simulations of the gravitational collapse of a collisionless system
of particles pre-dates the CDM model. Early simulations in the 1960's studied
the formation of elliptical galaxies from the collapse of a cold top-hat 
perturbation of stars \cite{1961AJ.....66..590V,1964AnAp...27...83H,1970AJ.....75...13P}.
The resulting virialisation process gave rise
to equilibrium structures with de Vaucouleurs \cite{1948AnAp...11..247D} or
Einasto \cite{1968PTarO..36..414E,1989A&A...223...89E} type density 
profiles, similar to observations of elliptical galaxies. 
It is remarkable that the end state of almost any gravitational collapse, independent of the small scale structure
and hierarchical merging pattern, leads to a similar global structure of the final equilibrium system \cite{1999ApJ...517...64H,Moore:1999gc,2006AJ....132.2685M}.

Computer simulations in the 70's attempted to follow the expansion and a collapse
of a spherical overdensity to relate to the observed properties of virialised structures such as galaxy
clusters \cite{1976MNRAS.177..717W}. Using a random distribution of particles with a Poisson power spectrum
lead to the initial formation of many bound clumps, however it was observed
that these bound structures were destroyed as the final system formed - resulting 
in a smooth distribution of matter. This overmerging problem persisted for over
two decades and motivated the development of semi-analytical models for galaxy formation \cite{1978MNRAS.183..341W}.

During the 1980's, it was proposed that cosmic structure formation
follows a dominant, non-baryonic cold dark matter (CDM) component \cite{1983ApJ...274....1P}.
Cold dark matter could consist of new and yet undiscovered weakly-interacting massive particles (WIMPs),
which occur for example in super-symmetric extensions of the Standard Model of particle physics \cite{1996PhR...267..195J}.
"Cold" means that these particles have rather small thermal velocities,
which allows the formation of very small structures, typically down to far below one solar mass
\cite{Green:2003un,2006PhRvL..97c1301P,2006PhRvD..74f3509B}. CDM together with the even more mysterious
dark energy (usually denoted "$\Lambda$") are the dominant components of the $\Lambda$CDM model, in which all the
ordinary matter accounts for only 4.6 percent of the total.
$\Lambda$CDM has by now become the "standard cosmological model" and its parameters 
(and therefore the initial conditions for structure formation)
are now known to a reasonable precision \cite{Dunkley:2008ie}.

Computer simulations allow to follow the non-linear evolution of perturbations,
starting from realistic and well constrained cosmological initial conditions.
The final quasi-equilibrium structures are the dark matter
halos that are observed to surround galaxies and galaxy clusters.
During the 1980's, the first simulations of the CDM model were carried out. 
Large cubes of the universe were simulated in an attempt to match
the large scale clustering of galaxies. Some of the most basic properties of collapsed
structures were discovered - the distribution of halo shapes, spin parameters etc \cite{1985Natur.317..595F,1986Natur.322..329Q}.
It was not until the simulations of Dubinski \& Carlberg  that 
individual objects were simulated at sufficiently high resolution to resolve their inner structure on scales that could be compared
with observations \cite{1991ApJ...378..496D}. Using a million particle simulation of a cluster mass halo run on a single 
workstation for an entire year, these authors found central cusps and density profiles 
with a continuously varying slope as a function
of radius. They fit Hernquist profiles to their initial simulations but an NFW profile \cite{1996ApJ...462..563N} provides an equally good fit (see Figure 1). 
Most likely due to a large softening length, the final virialised structure was almost completely smooth. 

\begin{figure}
\begin{center}
\includegraphics[scale=1.0]{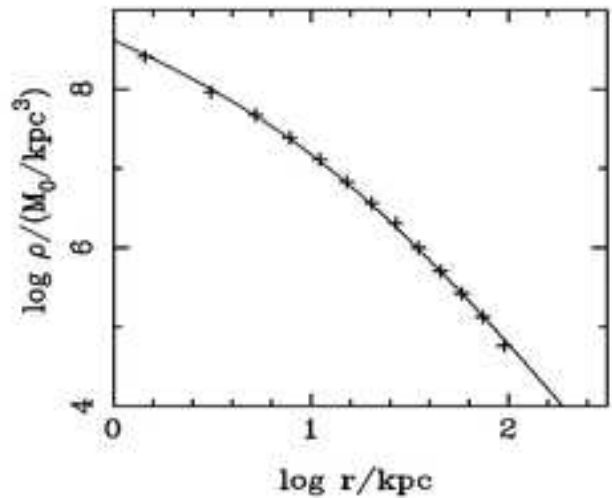}
\end{center}
\caption{Density profile of the million particle dark matter halo
simulation of Dubinski \& Carlberg 1990 (crosses).
The solid line shows the best fit NFW profile (Eqn. \ref{gnfw}) to the original data. This Figure was
adapted from \cite{1990ApJ...356..359H} by John Dubinski and it is reproduced
here with his permission.
}
\label{dubinski-nfw}
\end{figure}

Navarro et al. (1996) published results of simulations of halo density profiles from scales of galaxies to galaxy clusters.
They demonstrated that all halos could be reasonably well fit by a simple function (Eqn. \ref{gnfw}) with a concentration parameter
that was related to the halo mass \cite{1996ApJ...462..563N}. However, with less than $10^4$ particles only the mass profile beyond about 5-10 percent of the virial radius was resolved reliably. Shortly afterwards, simulations with $10^6$ particles showed cusps steeper than $r^{-1}$ down to their innermost resolved point near one percent of the virial radius \cite{Moore:1997sg}. These simulations also resolved the overmerging problem \cite{Moore:1995kj} - the resolution was sufficient to resolve cusps in the progenitor halos enabling the structures to survive the merging hierarchy \cite{Moore:1997sg,Ghigna:1998vn,Klypin:1999uc}. The final surviving substructure population is a relic of the entire merger history of a given CDM halo.

\subsection{State of the art simulations - convergence or discrepancies?}

Algorithmic and hardware development have increased the mass and spatial resolution 
by orders of magnitude (parallel computing, special 
purpose hardware, graphics pipelines etc). 
The first simulations used just a few hundred particles with
length resolution that was a large fraction of the final structure. Today we can simulate
individual collapsed structures, in a full cosmological context with up to $10^9$ particles
and spatial resolution that is better than $0.1\%$ of the virialised region and $\sim 10^5$
substructure halos can be resolved \cite{Diemand:2008in,Stadel:2008pn,Springel:2008cc}. 

The first billion particle halo simulation "Via Lactea II" (VL-II hereafter) \cite{Diemand:2008in}
was published 2008. About half a year later Springel et al. posted a preprint (0809.0898v1)
in which significant discrepancies between VL-II and their Aquarius simulations were claimed.
% just to later unveil essentially the same results as found earlier in VL-II. 
For some reasons these authors failed to acknowledge
that the Aquarius and VL-II simulations are perfectly consistent, which has caused some confusion.
% \footnote{Incidentally, these claims have later been taken back in the final, published version of this paper \cite{Springel:2008cc}.}.
Below we clarify the situation and to we hope to restore confidence in the robustness
of the current state-of-the-art simulations like VL-II\cite{Diemand:2008in}, GHALO\cite{Stadel:2008pn}
and Aquarius\cite{Springel:2008cc}.

Springel et al. 2008 claim two main discrepancies between VL-II and Aquarius:
subhalos in Aquarius are more concentrated than in VL-II and the abundance of subhalos as a function
of their peak circular velocity (see Section \ref{subhaloabundance}) is about 30 percent higher in Aquarius.
However, the Aquarius simulations still adopt the WMAP-1yr cosmological parameters,
especially $\sigma_8 = 0.9$, $n_s=1.0$ (see e.g. \cite{Dunkley:2008ie,Baumann:2008bn} for definitions and current constraints).
Both values are significantly higher (about 3 standard deviations) than the latest WMAP-5yr
values\cite{Dunkley:2008ie} and introduce much higher typical amplitudes for the small scale fluctuations
in the Aquarius initial conditions. VL-II on the other hand used the newer WMAP-3yr
values $\sigma_8 = 0.74$, $n_s=0.95$, which is somewhat lower, but
consistent (within one $\sigma$) with the latest parameters. The earlier VL-I simulation\cite{Diemand:2006ik}
had $\sigma_8 = 0.74$, $n_s\simeq0.9$, due to an error in the initial conditions generator
GRAFIC2\cite{Bertschinger:2001ng}, which leads to somewhat
lower (sub-)halo concentrations and velocity functions.

The substantially different
amplitudes of the typical initial small scale fluctuations have to be taken into account when the properties
of VL-II and Aquarius subhalos are compared. It is well known that halos and subhalos form earlier in the
WMAP-1yr cosmology, which leads to higher halo and subhalo concentrations \cite{Maccio':2008xb}. The
concentration difference between WMAP-1yr and 3-yr matches {\it exactly} with the "discrepancy" claimed
by Springel et al. 2008, i.e. the concentrations in VL-II and Aquarius differ by precisely the expected amount.
In other words, there is no discrepancy between the simulations, just a difference in adopted cosmological parameters.
The older cosmology used in Aquarius simply leads to significantly higher halo and subhalo concentrations compared to 
what's expected in a $\Lambda$CDM Universe with up-to-date parameters \cite{Maccio':2008xb}.

Higher subhalo concentrations also cause a higher subhalo peak circular velocity at the same subhalo
mass, i.e. a higher subhalo velocity function is expected in a cosmology
with too much small scale power \cite{Zentner:2003yd}. Therefore the higher subhalo velocity function, the second claimed 'discrepancy',
is also an expected consequence of the excess small scale fluctuations in the cosmology chosen for the Aquarius simulations.

Even if both simulations would have adopted the same cosmology,
a 30 percent difference in subhalo velocity functions of individual host halos would not be a
significant discrepancy, since velocity functions have substantial halo-to-halo scatter \cite{Ishiyama:2008xe}.
The small scatter found among the six pre-selected Aquarius halos cannot be generalised to the full population
of galaxy halos, which shows much larger variation \cite{Ishiyama:2008xe}.

After allowing for the differences caused by the different cosmologies the main results of Via Lactea and Aquarius, which used
different initial condition generators, different simulation codes and different analysis tools, are consistent.
This is a reassuring confirmation of the robustness and maturity of cosmological N-body simulations. Current simulations
are excellent approximations to cold dark matter halos on a large mass and spatial range. For most applications the
uncertainties in N-body results (e.g. on the very inner density profiles, see Section \ref{densityprofiles}) are already much smaller than the
expected, but still poorly constrained effects of galaxy formation
on the dark matter distribution, see Section \ref{baryons}.

\begin{figure*}
\includegraphics[scale=1.0]{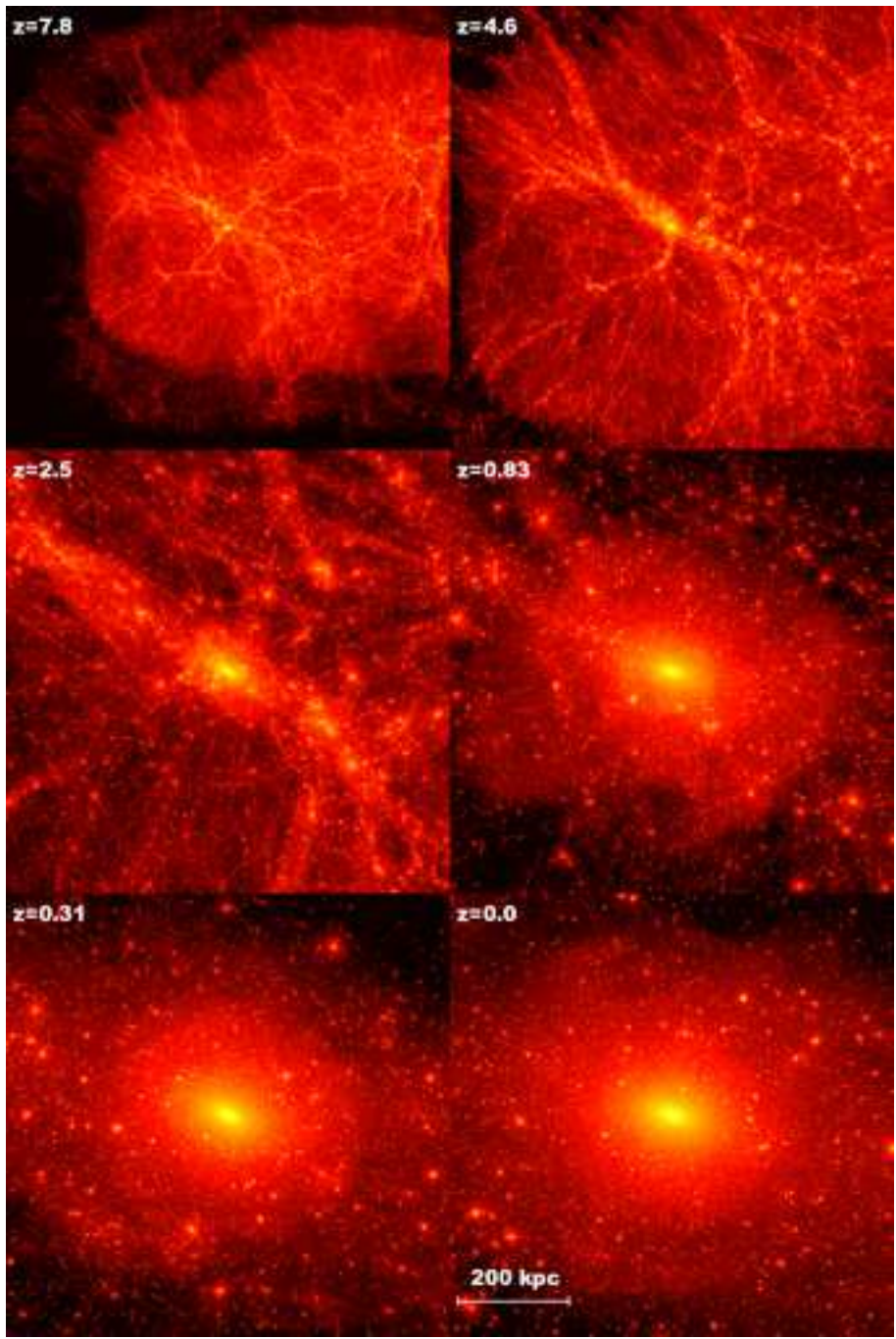}
\caption{Dark matter density maps from the ``Via Lactea II'' (VL-II) simulation \cite{Diemand:2008in}.
Cubes of 800 proper kpc are shown
at different redshifts, always centered on the main progenitor halo.
VL-II has a mass resolution of 4,100 $\msun$ and a 
force resolution of 40 pc. 
Initial conditions were generated with a modified, parallel version of GRAFIC2 \cite{Bertschinger:2001ng}.
The high resolution region (some of its border is visible in the upper panels)
is embedded within a large periodic box
(40 comoving Mpc) to account for the large scale tidal forces.
More images, movies and data are available at
http://www.ucolick.org/$\sim$diemand/vl/
}
\label{sixpanels}
\end{figure*}

\begin{figure}
\includegraphics[scale=1.0]{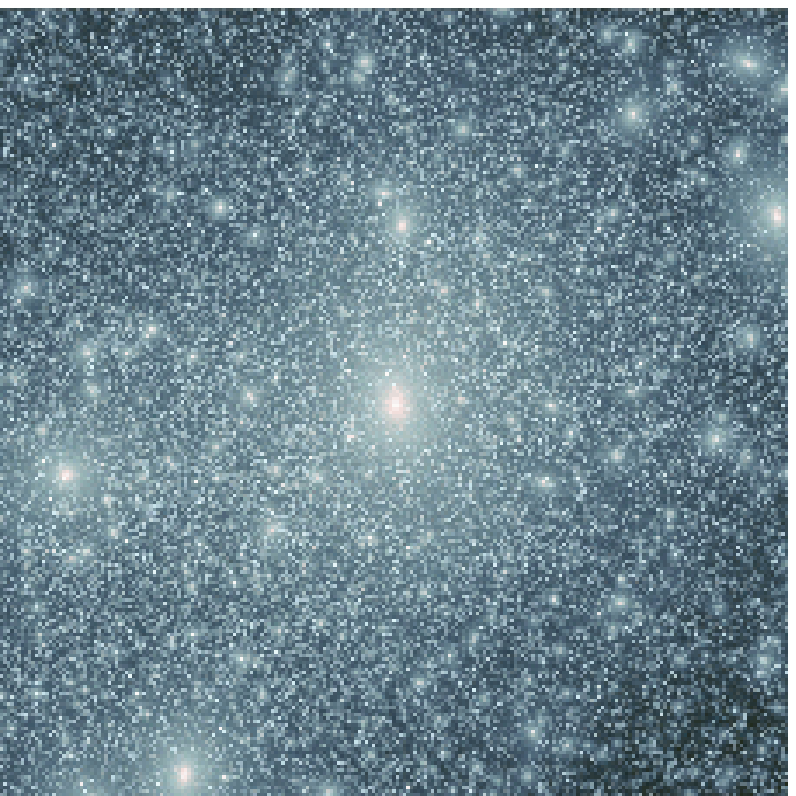}
\caption{Dark matter density map within the inner
200 kpc of the ``GHALO'' simulation \cite{Stadel:2008pn}.
This galaxy scale halo is resolved with over one billion particles of
1000 $\msun$ each. It contains over 100,000 subhalos, the largest are visible as bright
spots in this image.}
\label{ghalo}
\end{figure}

\section{Results from collisionless simulations}

Before we summarize the main simulation results on the properties of dark matter halos we introduce
definitions of the mass and radius of dark matter halos. Deciding which material belongs to a halo and
what lies beyond it is a non-trivial question, which has not received much attention until
quite recently \cite{Prada:2005mx,Diemand:2007qr,Cuesta:2007it,Diemand:2008gf}.

The usual halo definitions are still based on the simple spherical collapse picture
\cite{1975ApJ...201..296G,1977ApJ...218..592G}. These models assume that
there is no kinetic energy in an overdense sphere of matter as it turns around and that this sphere virializes, rather than
falling back radially into one point. The virial theorem predicts that the final halo radius is 0.5 of its turnaround radius
and that this "virial radius" would enclose an overdensity of $178$ times the critical density 
of the universe $\rho_{\rm crit}$ in an
Einstein-deSitter cosmology ($\Omega_M=1.0$). For the now favored $\Lambda$CDM cosmology
this overdensity is close to 100 times $\rho_{\rm crit}$ or roughly 300 times the mean matter density $\rho_{\rm M}$. However
CDM halos form very differently than assumed in the classical spherical collapse model:
material ending up near this so called "virial radius" did not undergo a collapse by anywhere near a factor of two \cite{Diemand:2007qr}.
The ratio of final radius to turnaround radius is much larger in the outer halo and typical orbits extend out to 90\%
of the turnaround radius \cite{Diemand:2008gf}, i.e. well beyond the formal virial radius. All the conventional
overdensity definitions, by which the halo radius encloses of the order of 200 times $\rho_{\rm crit}$ or $\rho_{\rm M}$,
are therefore too small, and they underestimate the extent and mass of dark matter halos significantly \cite{Prada:2005mx}.
However, spherical overdensities are easy to measure in simulations and useful for comparisons. In the following we adopt the largest of
these widely used halo radius definitions: $\rtwo$, defined to enclose 200 times $\rho_{\rm M}$. As a proxy for halo mass we will use 
$\mtwo = M(<\rtwo)$.

One has to keep in mind that practical, ad hoc halo definitions like $\rtwo$ and $\mtwo$ are rather different from the
concept of halo mass defined in theoretical models of structure formation. This can be confusing and has affected many attempts
for comparisons of theory with simulations, e.g.: halo mass functions (see \ref{subsec:massfunction});
halo mergers and merger trees are well defined in theory, but not in practice since the $\mtwo$ of the remnant
may be smaller than the sum of its progenitors $\mtwo$ and because some (sub-)halos do not
stay within $\rtwo$, i.e. they de-merge \cite{Moore:2004tga,Gill:2004vj,Diemand:2007qr,Diemand:2008gf};
by convention we will also refer only to halos within the virial radius of a larger halo
as "subhalos", but this actually miss-classifies many true satellite halos beyond this ad hoc scale as field halos,
which introduces environmental dependencies into the total "field" halo population \cite{Diemand:2007qr};
the halo mass $\mtwo(z)$ and radius $\rtwo(z)$ grow non-stop by definition, even when no mass is accreted \cite{Diemand:2007qr}.

\subsection{Mass function of halos}
\label{subsec:massfunction}

For a range of practical halo mass definitions, e.g. based on a given spherical overdensity, the abundance of field halos as a function of mass
and redshift can be accurately measured in cosmological simulations of large, representative, periodic volumes \cite{2007ApJ...671.1160L,Tinker:2008ff}.

The classic analytical approach by Press \& Schechter \cite{1974ApJ...187..425P} combines the statistics of a hierarchical, Gaussian random field with
the spherical, radial top-hat collapse model and it predicts the abundance of collapsed objects as a function of mass.
This idealized model matches the halo abundances measured in simulations surprisingly well, but it
predicts too many small halos ($M<M_*$, where $M_*$ is the mass of a typical, one-sigma halo forming now). This difference is
usually interpreted as being caused by the assumption of a spherical collapse. Allowing for an ellipsoidal collapse introduces
free parameters, which allows for a very accurate fit to the measured mass functions \cite{Sheth:2001dp,Tinker:2008ff}.

Alternatively, one can argue that the main difference between the model and the simulation is 
the collapse (or lack thereof): the infalling mass hardly loses any
energy, but typically orbits back out to 90\% of its turnaround radius \cite{Diemand:2008gf}, i.e. significantly beyond the "virial" radius.
In other words, one cannot expect to find the collapsed mass predicted by the Press-Schechter approach inside the
overdensity radii commonly used to define halo masses in simulations. A recently suggested improvement is to use the stationary mass,
i.e. the mass within the largest radius with zero mean radial velocity \cite{Prada:2005mx,Cuesta:2007it}. The static mass exceeds the
"virial" mass substantially in halos below $M_*$ and at z=0 the Press-Schechter mass function
fits the abundance of halos as a function of their {\em static} mass very well \cite{Cuesta:2007it}. At z=1 and z=2 the
agreement is less impressive, the Press-Schechter mass function lies above simulated abundances \cite{Cuesta:2007it}.

Further improvements in these directions seem possible, since Press-Schechter predicts the collapsed mass, not the static
mass. Above $M_*$ accretion typically sets in near $r_{\rm vir}$, i.e. static and virial radius are similar \cite{Prada:2005mx}.
But also above $M_*$significant amounts of previously collapsed material are found beyond the virial
radius\cite{Moore:2004tga,Gill:2004vj,Diemand:2007qr,Diemand:2008gf}, so both the static and the virial mass seem to underestimate
the collapsed mass for halos above $M_*$.

\subsection{Global halo properties}

Here we briefly summarize some basic, global properties of CDM halos:
concentration and density profiles; shapes, spin and velocity distribution.

\subsubsection{Halo formation, density profiles and concentrations}\label{concentrations}

Self-similar infall models predict scale free, nearly isothermal
($\rho \propto r^{-2}$) profiles
\cite{1975ApJ...201..296G,1977ApJ...218..592G,1984ApJ...281....1F,1985ApJS...58...39B}.
Simulated profiles are found to be steeper than $\rho \propto
r^{-2}$ in the outer parts and shallower in the inner regions
\citep{1991ApJ...378..496D}, i.e. their circular velocity profiles $V_c(r) = \sqrt{G M(<r) / r}$ have 
a well defined peak, which serves as a natural halo size scale $\vmax$ and scale radius $\rvmax$.

Self similar spherical infall maintains a distinct relation between the initial conditions and the final structure,
while violent relaxation assumes information on the initial conditions is chaotically lost.
Simulations show that despite their non-isothermal density profiles, CDM halos
still do have much in common with idealized spherical infall models:
i) particles that collapse in rare, early peaks 
(e.g. formation sites of first stars and old globular clusters)
are located closer to the potential minima of the entire turnaround region due to peaks biasing and those particles do
end up closer to the center of the final halo \cite{Diemand:2005rd,Moore:2005jj}.
ii) the typical particle apocenter distances are close to their turnaround radii \cite{Diemand:2008gf}.
Modified infall models are indeed able to reproduce some of the features of halo density profiles found
in cosmological simulations \cite{2007MNRAS.376..393A}.

Over a mass range spanning 20 decades, from micro-halos to galaxy clusters, the spherically averaged CDM halo density profile
can be approximated with the same universal form (NFW \citep{1996ApJ...462..563N}):
\begin{equation}\label{gnfw}
\rho(r) = \frac{\rho_s}{(r/r_s)^{\gamma} (1 + r/r_s)^{3-\gamma}},
\end{equation}
where $\gamma =1$. The scale radius $r_s$ is related to the peak circular velocity scale
by $r_{\rm Vmax} = 2.163 \; r_s$ and it is used to define the halo concentration
$c_{\rm vir} = r_{\rm vir} / r_s$, where the
"virial" radius $r_{\rm vir}$ is defined following one of the ad hoc overdensity criteria described above.
Halo concentrations $c_{\rm vir}$ (and equivalently scale densities $\rho_s$) are related to the halo
formation time: early forming halos tend to have higher $c_{\rm vir}$ and $\rho_s$ at z=0. In CDM the
typical amplitudes of density fluctuations $\sigma(M)$ decreases from dwarf to galaxy cluster scales, i.e. 
smaller halos form earlier on average than larger ones and they 
end up having higher median concentrations
\cite{1996ApJ...462..563N,Bullock:1999he,Kuhlen:2004rw,Maccio':2006nu,Maccio':2008xb}. 
At a given mass, the concentrations of individual halos have a large scatter:
the standard deviation in log $c_{\rm vir}$ is 0.18. On subsolar mass scales the CDM power
spectrum approaches $P(k) \propto k^{-3}$, i.e. $\sigma(M)$ approaches a constant, which
leads to very similar halo formation times and halo concentrations over a wide range of
masses \cite{Green:2003un,Diemand:2005vz,Colafrancesco:2005ji}.
Even Earth mass micro-halos, the first and smallest systems in the CDM hierarchy,
have NFW-like density profiles. A systematic study of their typical concentrations is still lacking,
but values found in the small sample of \cite{Diemand:2005vz}
($c_{\rm vir}(z=0) \simeq 80$) seem consistent with the 
predictions of the Bullock et al. model \cite{Colafrancesco:2005ji}.

A simpler and more general measure of halo concentrations is the mean density within $\rvmax$.
It is well defined both for isolated halos {\it and} subhalos and it is independent of assumptions on their
"virial" radius or their density profile \cite{Diemand:2007qr}:
\begin{equation}
c_{V} \equiv \frac{\bar{\rho}(<r_{\rm Vmax})}{\rho_{\rm crit,0}} 
% = 2 \left( \frac{V_{\rm max}}{H_0 r_{\rm Vmax}} \right)^2
=  \left( \frac{V_{\rm max} }{ r_{\rm Vmax}} \right)^2 \frac{3}{4\pi G \rho_{\rm crit,0}} \; .
\end{equation}
For an NFW halo it is easy too convert
from $c_{V}$ to $c_{\rm vir}$ [\cite{Diemand:2007qr}]. Since the NFW form is not a very good fit to most
CDM halos, the measured $c_{\rm vir}$ depend somewhat on the details of the fitting procedure
\cite{Bullock:1999he,Kuhlen:2004rw,Maccio':2006nu,Maccio':2008xb}.
These complications could be avoided by using $c_{V}$, which is a robust concentration
measure as long as $\vmax$ and $\rvmax$ are resolved in the simulation.

\subsubsection{Inner density profiles}\label{densityprofiles}

The NFW form (Eqn. \ref{gnfw}) was proposed as a fit to CDM density profiles in the radial range from
0.01 $\rtwo$ to $\rtwo$ [\cite{1996ApJ...462..563N}].
As larger simulations started to resolve scales around 0.01 $\rtwo$ it became clear that most halos
are significantly denser in their inner parts than their best fit NFW profile
\cite{Moore:1997sg,Ghigna:1998vn,Ghigna:1999sn,Klypin:2000hk,Fukushige:2000ar,Fukushige:2001hs,Fukushige:2003xc,
Tasitsiomi:2003dd,Navarro:2003ew,Diemand:2004wh}.
Samples of high resolution, relaxed, isolated CDM halos also
revealed significant halo-to-halo scatter \cite{Navarro:2003ew,Diemand:2004wh}:
some halos follow the NFW form quite well, while others are better approximated with a steeper profile as
suggested by \cite{Moore:1997sg}. Most halos lie somewhere in between and their average deviations from both fitting
functions are larger than 10 percent (more than 20 percent at some radii) \cite{Diemand:2004wh}.
Improved fits require an additional free parameter to account for the substantial halo-to-halo scatter. Good fits
are obtained by letting the inner slope $\gamma$ of the NFW function (Eqn. \ref{gnfw}) become a free parameter, instead of
forcing $\gamma=1$. Another option is to use the Einasto profile \cite{1968PTarO..36..414E,1989A&A...223...89E,Navarro:2003ew}
\begin{equation}\label{einasto}
\rho(r) = \rho_s  e^{-\frac{2}{\alpha} \left[ (r/r_{s})^{\alpha} - 1 \right] } \; ,
\end{equation}
where $\alpha$ is the additional free parameter. Down to 0.03 $\rvmax$ both forms are very similar and both
fit simulated CDM density profiles very well \cite{Navarro:2003ew,Diemand:2004wh,2006AJ....132.2685M,Graham:2006ae}. The two functions
only differ significantly at very small radii (below about 0.6 percent of $\rvmax$, i.e. below about 500 pc in a galaxy halo).
Only very few (if any) simulations are currently able to resolve such small scales. 
The halos in \cite{Diemand:2005wv,Diemand:2008in} are denser than their Einasto fits in the inner parts and they
are well approximated by $\gamma \simeq 1.2$ cusps. The same is true for the billion particle GHALO down to
400 pc \cite{Stadel:2008pn}. However, the higher mass resolution of GHALO might allow significantly smaller scales to be resolved and a 
new functional form for the fitting function of the density profile has been proposed \cite{Stadel:2008pn}:
\begin{equation}
\label{eqn:lambda}
\rho(r) = \rho_0 e^{-\lambda\left[\ln(1 + r/R_{\lambda}) \right] ^2}.
\end{equation}
This function has a constant logarithmic slope down to a scale $R_{\lambda}$, beyond 
which it approaches the central maximum density $\rho_0$ as $r \rightarrow 0$. If 
one makes a plot of $d\ln\rho/d\ln(1+r/R_{\lambda})$ vs. $\ln(1+r/R_{\lambda})$, then this profile
forms an exact straight line with slope $-2\lambda$. Good fits are obtained with $\lambda \simeq 0.10$ \cite{Stadel:2008pn}.

However, to demonstrate convergence on such extremely small scales, several issues have to be addressed first:
One necessary condition is to resolve the short dynamical times in these inner, very high density regions. 
The widely used, but unphysical, time step criterion $\Delta t\propto \sqrt{\epsilon/|a|}$, where $\epsilon$ is the force softening, $|a|$ the local acceleration,
can lead to artificially low central densities \cite{Diemand:2005wv}.
The VL-II and GHALO simulations use an improved criterion based on the local dynamical time \cite{Zemp:2006vh}.
Another caveat is that the material ending up in the inner 500 pc of a galaxy halo comes mostly from very early forming,
dense progenitor halos ($4 \sigma$-peaks and higher \cite{Diemand:2005rd}). At the starting redshifts of these simulations
the variation on the smallest resolved mass scales $\sigma(M_{\rm min})$, is in the range 0.1 to 0.2.
$4 \sigma$-peaks already have the (mildly) nonlinear density contrast of 0.4 to 0.8,
which artificially lowers their formation redshift and scale density. The common practice of
starting the simulations when $\sigma(M_{\rm min}) \simeq$ 0.1 artificially lowers
the halo abundace at high redshifts \cite{2007ApJ...671.1160L}.
Furthermore, relaxation effects will be present within these first structures to collapse since they
are always resolved by a small number of particles \cite{Binney:2001fi,Diemand:2003uv}.

\begin{figure}
\includegraphics[scale=0.4]{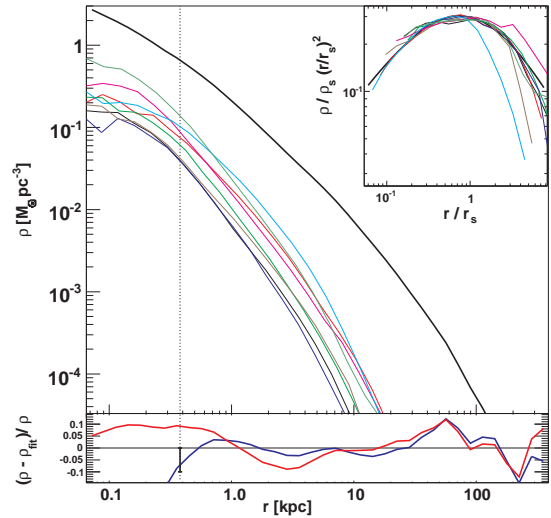}
\caption[]{Density profiles of Via Lactea halo and several subhalos.
\small {\it Main panel}: Profile of the main cold dark matter
halo ({\it thick line}) and of eight large subhalos ({\it thin lines}).
The {\it lower panel} gives the relative differences between
the simulated main halo profile and the Einasto fitting formula (Eqn. \ref{einasto}),
with best fit parameters: $\alpha =0.170$, $r_s = 21.5\, {\rm kpc}$,     
$\rho_s=1.73 \times 10^{-3} \sden$
({\it red curve}) and function (\ref{gnfw}) 
with a best fit inner slope of $\gamma = 1.24$,
$r_s = 28.1\, {\rm kpc}$, $\rho_s=3.50 \times 10^{-3} \sden$ ({\it blue curve}).
The vertical dotted line indicates the
estimated convergence radius of 380 pc: simulated local densities
are only lower limits inside of 380 pc and they
should be correct to within 10\% outside this region.
The cuspy profile is a good fit to the inner halo, while the Einasto profile
has a too shallow slope in the inner few kpc, causing it to overestimate
densities around 4 kpc and to underestimate them
at all radii smaller than 1 kpc. The same behavior in the inner few kpc is found at also at higher
redshifts, while the large residuals in the outer halos on the other hand are transient
features. {\it Inset}: Rescaled host (thick line) and subhalo (thin lines) density profiles multiplied
by radius square to reduce the vertical range of the figure.
This figure was reproduced from \cite{Diemand:2008in}.
}
\label{denpros}
\end{figure}

Beyond 0.6 percent of $\rvmax$ there is a general consensus that halos are denser than
NFW and that they are well fitted by functions which are steeper and cuspier than NFW at these radii
like Eqn. \ref{gnfw} with $\gamma \simeq 1.2$, or the Einasto form (Eqn. \ref{einasto}) with $\gamma \simeq 0.17$. 
To resolve the remaining uncertainties in the density profiles of pure CDM halos
below 0.6 percent of $\rvmax$ is of some numerical and theoretical interest,
but it does not affect the observational predictions from CDM significantly:
\begin{itemize}
\item In galaxies, groups and clusters these scales are dominated by the baryons. Predictions on the dark matter and total
mass distribution require a realistic treatment of the baryons and their dynamical interactions with the dark matter (see Section \ref{baryons}).
\item The predicted kinematics of dark matter dominated dwarf galaxies are indistinguishable for these two density profiles, even if
a large numbers of accurate proper motion measurements become available \cite{Strigari:2007vn}.
\item The annihilation properties of small dark matter dominated halos (see Section \ref{sec:indirectdetection})
do not change significantly: The resulting annihilation luminosities of both fits and of the measured profile
in Figure \ref{denpros} all lie within 5 percent and the half light radii are within 10 percent of each other. The best NFW fit to the VL-II
density profile however underestimates the halo luminosity by a factor of 1.4 and it overestimates the half light radius by 1.3.
\end{itemize}
 
\subsubsection{Shapes, spin and velocity distributions}\label{shapes}

Another main result from cosmological collision-less simulations is that CDM halos are rather elongated,
significantly more than for example elliptical galaxies \cite{1991ApJ...378..496D}.
The mean minor-to-major axis ratio is well described by  
$<s> = 0.54 (M_{\rm vir}/M_*)^{-0.050}$, where s is measured at 0.3 $r_{\rm vir}$.
The rms-scatter around the mean is about $0.1$ [\cite{Allgood:2005eu}]. The length of the intermediate axis
is usually closer to the minor axis, i.e. most halos are prolate. Towards the center CDM halos
are even more elongated, i.e. the axis ratios become smaller while their orientations remain fairly well aligned.
At a growing radius like $0.3 r_{\rm vir}$, halo samples become rounder with time \cite{Jing:2002np,Allgood:2005eu},
while individual halos have quite stable shapes at some fixed, inner radius (except of course during major mergers) \cite{Kuhlen:2007ku}.

CDM halos are supported by nearly isotropic velocity dispersions, not by rotation.
The tangential and radial velocity dispersions measured in spherical bins define
the anisotropy parameter $\beta = 1 - 0.5 \sigma_{\rm tangential}^2 / \sigma_{\rm radial}^2$.
The typical, relaxed CDM halo has $\beta(r) \simeq 0.35 (r/r_{\rm vir})^{0.35}$, i.e. $\beta$ changes
from nearly isotropic near the center to mildly radial in the outer halo \cite{Diemand:2004kx,Wojtak:2005xq}.

The velocity distributions of CDM halos are Gaussian only near the scale radius, i.e. where the density profile
is roughly isothermal ($\rho \propto r^{-2}$). Analytical calculations predict more peaked distributions for the inner
halo, and broader distributions beyond the scale radius \cite{1990ApJ...356..359H,Kazantzidis:2003im,Wojtak:2005xq,Wojtak:2008mg}
in agreement with the measured distributions in CDM simulations \cite{Hansen:2005yj,Wojtak:2005xq,Wojtak:2008mg,Fairbairn:2008gz}.

Halo shapes are supported by velocity dispersion ellipsoids, which are elongated in the same direction as the mass distribution, both
globally \cite{Allgood:2005eu} and also locally \cite{Zemp2008}. Fairly symmetric velocity distributions in the tangential directions are found
throughout the halo \cite{Zemp2008}, i.e. there is about as much negative as positive angular momentum material
relative to any given reference axis. Small asymmetries lead to a some residual
spin $\lambda = |\vec{J}| |E|^{1/2} / (GM^{5/2})$, which has a median of only $\bar{\lambda} = 0.04$ and a lognormal distribution of width $\sigma_{\lambda} \simeq 0.56$,
both for the dark matter and for adiabatic gas. However, the two spins are often poorly aligned, with a median misalignment
angle of 30 degrees \cite{vandenBosch:2002kw}. The dark matter angular momentum tends to align roughly with the minor
axis of the halo shape, with a mean misalignment of 25 degrees \cite{Bailin:2004wu}, while disk galaxy orientation and halo
shape beyond 0.1 $r_{\rm vir}$ are completely uncorrelated \cite{Bailin:2005xq}. It is often assumed that the
net halo angular momentum correlates with disk galaxy size and orientation, however only some
selected fraction of the halo material
with its wide variety of angular momenta can be incorporated into 
a realistic disk \cite{vandenBosch:2002kw}. In our Galaxy for example, a significant fraction
of the total available baryonic angular momentum is not in the disk but in the polar orbit of the Magellanic clouds
given their relatively large distances, masses and proper motions \cite{Kallivayalil:2006ry}.
How exactly disk galaxies form out the angular momentum distributions available in $\Lambda$CDM halos remains an open question
despite much recent progress \cite{Governato:2006cq,Mayer:2008mr}.

\subsection{Substructure}

A major contribution of N-body simulations to our understanding of structure formation
was to demonstrate how hierarchical merging gives rise to a vast amount of
surviving substructure, both gravitationally bound (subalos) and unbound (streams).

\subsubsection{Subhalo abundance: velocity and mass functions}\label{subhaloabundance}

\begin{figure}
\begin{center}
\includegraphics[scale=0.4]{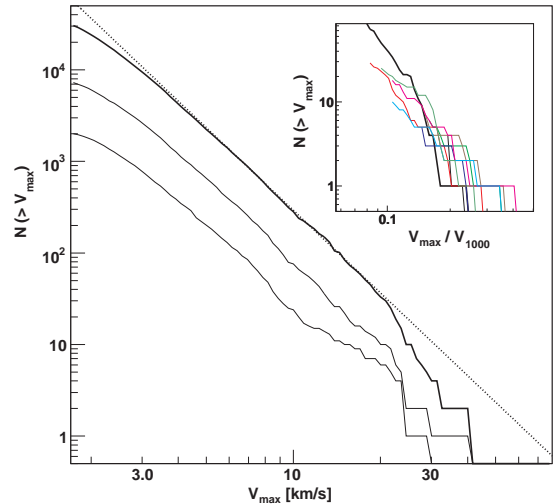}
\end{center}
\caption{Subhalo and sub-subhalo abundances in the VL-II halo. Number of subhalos above
$\vmax$ within $\rtwo=402$ kpc (thick solid lines) and within 100 and 50 kpc of the
galactic center (thin solid lines). The dotted line is
$N(>\vmax)=0.036\,(\vmax / V_{\rm max,host})^{-3}$, where $V_{\rm max,host} = 201 \kms$
(at $r_{\rm Vmax,host}= 60$ kpc).
It fits the subhalo abundance above $\vmax \simeq 3.5 \kms$.
The number of smaller subhalos is artificially reduced by numerical
limitations. The {\it inset} shows the sub-subhalo abundance within $r_{1000}$ 
(enclosing 1000 times the mean matter density) of the centers
of eight (same ones as in Fig. \ref{denpros}) large subhalos (thin solid lines). $r_{1000}$
is well inside of the tidal radius for these systems. The thick solid line
shows the subhalo abundace of the host halo inside of its $r_{1000} = 213$ kpc.
The (sub-)subhalo $\vmax$ values are given in
units of $V_{1000}=\sqrt{GM(<r_{1000}) / r_{1000} }$
of the corresponding host (sub-)halo. Lines stop at $\vmax = 2 \kms$.
The mean sub-substructure abundance is consistent with 
the scaled down version of main halo, and both the mean abundance and the
scatter are similar to the values found in \cite{Reed:2004vg,Ishiyama:2008xe} for distinct {\it field} halos.
This figure is from \cite{Diemand:2008in}. 
}
\label{velf}
\end{figure}

The abundance of subhalos within $\rtwo=402$ kpc of in the VL-II simulation (Figure \ref{velf}) is well approximated by
\begin{equation} \label{Nvmax}
N(>\vmax)=0.036\,(\vmax / V_{\rm max,host})^{-3} \;, 
\end{equation}
where $V_{\rm max,host} = 201 \kms$. It is close to the median abundance found in a large sample of halos simulated
using similar cosmological parameters \cite{Ishiyama:2008xe}. In earlier halo samples using a higher normalization
of the power spectrum ($\sigma_8=1.0$) a higher normalization of 0.042 for the median abundance was found \cite{Reed:2004vg}.
This difference is consistent with the expected cosmology dependence \cite{Zentner:2003yd}. Both samples demonstrate
significant halo-to-halo scatter of about a factor of two. Some of the variation comes from the scatter in halo 
concentration, which introduces scatter in $V_{\rm max,host}$ at a given host halo mass. 
Normalizing the subhalo function to $V_c(\rtwo)$ (i.e. to $\mtwo$) reduces the scatter to about a factor of 1.25 [\cite{Ishiyama:2008xe}[. 
Without normalizing to the host halo size, the differential $\vmax$-distribution function of subhalos in dwarf galaxies to cluster halos scatters around
\begin{equation} \label{dnvmax}
dn/d\vmax = 1.5 \, \times 10^8 \vmax^{-4.5} \, (h^3 {\rm Mpc}^{-3}  \kms )\;, 
\end{equation}
again with a halo-to-halo scatter of about a factor of two \cite{Reed:2004vg}.
The median subhalo abundance given by the normalized (\ref{Nvmax}) or non-normalized
velocity function (\ref{dnvmax}) seems to be approximately self-similar,
i.e. independent of host halo mass and redshift \cite{Reed:2004vg}.
Even tiny host halos ($< \msun$) show similar abundances, perhaps surprisingly so given that the density contrasts between subhalos and the host
are much smaller on these scales due to their similar formation times caused by the nearly flat $\sigma(M)$ 
\cite{Diemand:2006ey,Elahi:2008cd}.

While cosmological simulations are able to resolve the subhalo content of a given dark matter halo accurately, the
exact abundance of substructure around a given Galaxy, like for example the Milky Way, remains uncertain:
\begin{itemize}
\item The $V_{\rm max,host}$ of the CDM halo in which a galaxy with a rotation speed of 
220 $\kms$ would form could lie anywhere between 160\--220$\kms$ [\cite{Klypin:2001xu,Dutton:2006vi,Smith:2006ym}].
The resulting uncertainty in $N(>\vmax)$ spans about a factor of $(220/160)^3 = 2.6$.
\item At a given $V_{\rm max,host}$, and also at a given $\mtwo$, the subhalo abundance within $\rtwo$
has a substantial halo-to-halo scatter \cite{Reed:2004vg,Ishiyama:2008xe}.
\end{itemize}
These theoretical uncertainties have to be considered when CDM subhalos are compared to the subhalos around the
Milky Way satellite galaxies.

Cumulative mass functions of subhalos can be approximated by power-laws of the form $M^{-\alpha}$, with $\alpha = 1.9$ to 2.0
and normalized so that the mass in subhalos larger than $10^{-6} M_{\rm host}$ is between about 5 to 15 percent
\cite{Moore:1997sg,Ghigna:1998vn,Moore:1999nt,Klypin:1999uc,Ghigna:1999sn,Moore:2001vq,Kravtsov:2003sg,
DeLucia:2003xe,Kravtsov:2003sg,Diemand:2004kx,Gao:2004au,Diemand:2006ey,Diemand:2006ik,Diemand:2007qr,Madau:2008fr,Springel:2008cc}.
The steep slope of the mass function means that there is a significant amount of mass in small subhalos, which are
still unresolved in current simulations. Numerical convergence studies show that about four hundred particles
per subhalo are required to resolve the mass function within $\rtwo$ [\cite{Diemand:2004kx}].
The convergence test in Springel et al. (2008) confirm these results, however subhalos
resolved with only 60 particles are still included in their mass
function\symbolfootnote[1]{In the dense inner regions of halos and subhalos the numerical requirement are actually
even higher\cite{Diemand:2004kx,Springel:2008cc}.
In these regions however Springel et al. \cite{Springel:2008cc}
include systems resolved with only 20 particles in their sub-substructure abundance estimates.},
which artificially lowers the slope $\alpha$ of their best fitting power-law.

\subsubsection{Subhalo evolution and their final spatial distribution}

\begin{figure}
\includegraphics[scale=0.4]{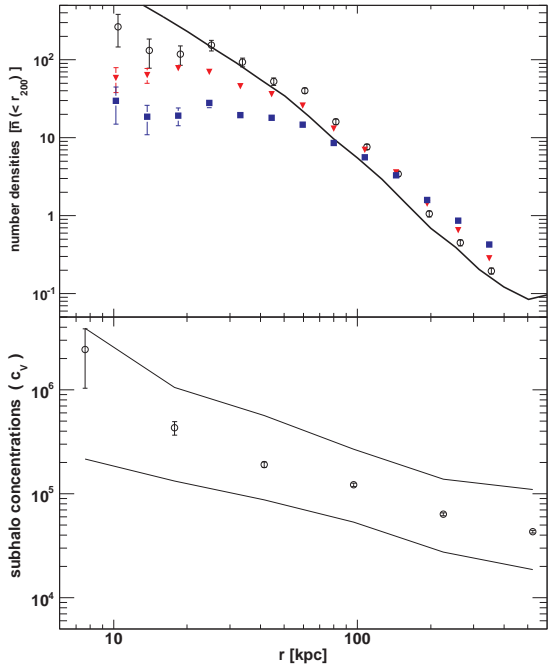}
\caption{Abundance and concentrations of subhalos vs. distance from the galactic center
in the VL-II simulation.
{\it Top:} The number density profile of subhalos (symbols) is more extended than  the dark matter
density profile $\rho(r)$ (thick line). The difference relative to the matter distribution is
largest for mass selected subhalo samples: Squares show a subhalo sample with a mass above
$8 \times 10^5 \msun$. Selecting subhalos by their peak circular velocity $\vmax$ reduces the difference:
Triangles are for subhalos with $\vmax > 3 \kms$. Selecting by subhalo annihilation luminosity
($L\propto \vmax^4 / \rvmax > (5 \kms)^4/(0.15 \kpc)$)
eliminates the difference throughout most of the halo (circles).
Subhalo luminosity closely traces the dark matter distribution,
except in the very inner regions (about 5\% of $r_{200}$, or 20 kpc for the VL-II halo).
{\it Bottom:} Subhalo concentrations $c_{V}$ (median and 68\% range are shown) 
increase towards the center, where the stronger tidal force remove more of the outer, low density
parts from the subhalos. 
To make sure their $c_{V}$ are resolved, only subhalos larger than $\vmax = 5 \kms$ are used.
The error bars indicate the statistical uncertainties in both panels. This figure was adapted from \cite{Diemand:2008in}. 
}
\label{radial}
\end{figure}

The abundance of field halos of a given (moderate) mass is proportional to the dark matter density of the environment.
This proportionality is altered as matter and halos fall into a host halo and tidal forces reduce the mass of systems,
especially those orbiting close to the host halo center. Subhalos selected by their final remaining mass are
therefore more extended than the matter distribution (see Figure \ref{radial})
\cite{Ghigna:1998vn,Kravtsov:1998vt,Colin:1999bh,Diemand:2004kx,Gao:2004au,Faltenbacher:2006rb,Diemand:2006ik,Diemand:2008in,Springel:2008cc}.
The number density of subhalos is independent of the mass threshold and roughly proportional to
the dark density times radius: $n_{sub,M0}(r) \propto \rho_{DM}(r) \times r$
 [\cite{Diemand:2004kx,Gao:2004au,Diemand:2007qr,Springel:2008cc}].
$\vmax$ is less affected by tides \cite{Kravtsov:2004cm,Diemand:2007qr} and $\vmax$ selected subhalos
are show less of a difference relative to the dark matter distribution (see Figure \ref{radial}). 

During tidal mass loss $\rvmax$ becomes smaller and the enclosed mean subhalo density $c_{V}$
increases. Subhalos near the center of the host halo end up having much larger concentrations
$c_{V}$ than field halos (see Figure \ref{radial}) \cite{Diemand:2007qr,Diemand:2008in}.

In terms of a subhalos dark matter annihilation luminosity 
$L \propto \vmax^4 / \rvmax \propto \vmax^3 \sqrt{c_{V}} $ (see Section \ref{sec:indirectdetection}), the increase in $c_{V}$
partially compensates for the reduced $\vmax$. 
The total annihilation luminosity
from the entire CDM subhalo population therefore traces the dark matter distribution,
even tough the mass in subhalos is much more extended (see Figure \ref{radial}).
In other words the total subhalo luminosity mostly (except close to the halo center)
follows the dark matter distribution, because
$L$ is not significantly affected by tidal stripping for most subhalos. 
This is not surprising, given that
half of a subhalos annihilation originates from the rather small radius of about 0.07 $\rvmax$.
Smaller halos than
those resolved in current simulations have higher typical concentrations, i.e. higher densities inside
0.07 $\rvmax$ and their annihilation luminosity is expected to resist tidal losses even better.

Tidal stripping removes mass from the outer, loosely bound regions of subhalos
\cite{Ghigna:1998vn,Kravtsov:2004cm,Diemand:2006ik}. Stripping is well approximated by removing
the mass beyond the tidal radius over some timescale \cite{2001ApJ...559..716T,Zentner:2003yd,Diemand:2007qr}. 
The tidal radius is defined so that the host halo density is
equal (or similar to) the subhalo density at the subhalos tidal radius. High density parts
of subhalos are therefore able to survive intact even close to the center of the host.
This explains why practically all (97\% since z=1) subhalos survive
until the present time, despite substantial mass loss in some cases \cite{Diemand:2007qr}.
Subhalos have an inner, tightly bound region unaffected by mass loss, whose extent depends on the
subhalo concentration and on its orbit \cite{Kazantzidis:2003hb,2006MNRAS.367..387R,Diemand:2006ik,Choi:2008xi}. The mass profile and the
substructure content deep inside subhalos are the same as found in the inner parts of field halos \cite{Diemand:2008in}.
These two findings are related, because tidal stripping removes smooth {\it and} clumpy material
without preference for one or the other, but from the outer parts only. The inner regions of subhalos
retain their cuspy density profiles (Figure \ref{denpros}) \cite{Kazantzidis:2003hb,Diemand:2006ik,Diemand:2008in,Springel:2008cc,Choi:2008xi}
{\it and} their substructure (Figure \ref{velf}) \cite{Diemand:2008in}.
The impression that subhalos should have shallower inner density profiles \cite{Stoehr:2002ht}
and less substructure\cite{Springel:2008cc} than the inner parts of field halos is caused by insufficient
numerical resolution.

Subhalos move on rather radial orbits. The median peri- to apocenter ratio is
1:6 [\cite{Ghigna:1998vn,Diemand:2007qr}], and only 5\% of the orbits are rounder than 2:3 [\cite{Diemand:2007qr}].
Only a few subhalos are massive enough to suffer significant dynamical friction, which causes 
decaying orbits, disproportionally large mass loss and even complete merging with the center
of the main halo in some cases \cite{Diemand:2006ik}.
Most subhalos, and dark matter particles, move on fairly regular orbits:
They reach nearly constant median apocentric distances, which lie
close to their initial turnaround radii \cite{Diemand:2008gf}. 
A few subhalos even gain energy during their pericenter passage in three-body interactions involving a larger
subhalo and the host halo \cite{Sales:2007hr}.

\subsubsection{Subhalo shapes and orientations}

The shapes of subhalos are similar to those of field halos (see \ref{shapes}), but subhalos
tend to be a bit rounder, especially the ones near the host halo center \cite{Kuhlen:2007ku}. Tidal interactions make individual subhalos
rounder over time \cite{Moore:2003eq}, and they also tend to align their major axis towards the center of the host
halo \cite{Kuhlen:2007ku,Faltenbacher:2007yw,Pereira:2007sw,Knebe:2008th}. The alignment is often maintained
over most of the subhalos orbits, except during pericenter passages \cite{Kuhlen:2007ku}. A similar 
radial alignment has been found for red galaxies in SDSS groups \cite{Faltenbacher:2007sk}. The major axes of the
Milky Way dwarf satellites might also be preferentially aligned radially, i.e. roughly towards us. If that is case
current mass estimates based on spherical models
(e.g. \cite{2007ApJ...670..313S,2007MNRAS.380..281M,Strigari:2007ma,2008Natur.454.1096S})
would be biased towards somewhat higher values.

\subsubsection{Other halo substructure: caustics, streams and voids}

\begin{figure}
\includegraphics[scale=1.0]{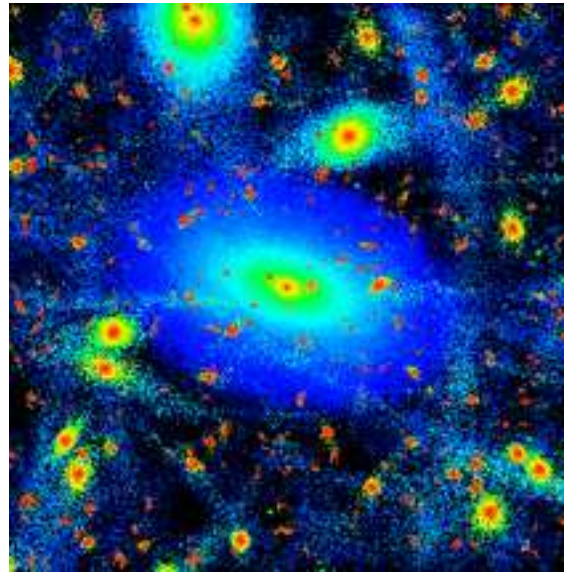}
\caption{Local phase space densities calculated with EnBiD \cite{Sharma:2005ih} 
in a 40 kpc cube in the center of the VL-II halo at z=0
(compare to the last panel in Figure \ref{sixpanels}). Subhalos
have very high central phase-space densities ($> 10^{-5}\psden$)
due to their steep inner density cusps and
their relatively small internal velocity dispersions.
The most clearly visible streams still have quite low densities 
(about 100 times below the local density) but owing to their low velocity dispersion
(about 10 times smaller than that of background particles) they just barely manage
to stand out in local phase space density (these streams have about $10^{-9}\psden$).
}
\label{psd}
\end{figure}

Besides the gravitationally bound, dense subhalos discussed above, there exists additional structure
in the phase space of CDM halos. Current simulations are now starting to resolve some
of this structure (see Figure \ref{psd}), although finite mass resolution and artificial numerical heating
\cite{Binney:2001fi,Diemand:2003uv}
still severely limits our ability to detect and resolve fine grained phase space structure. The coherent elongated features in
Figure \ref{psd} are dark matter streams which form out of material removed from accreted and disrupted subhalos.
In cases where the disrupted subhalo hosted a luminous satellite galaxy, the resulting
streams would contain not only dark matter but also stars and produce detectable features in galactic stellar
halos \cite{2005ApJ...635..931B}. This process explains the origin of
stellar streams observed around the Milky Way \cite{2006ApJ...642L.137B,2008A&ARv..15..145H}.

Cosmological infall into a halo does at some times lead to the characteristic patterns in the $f(v_r,r)$ plane \cite{Diemand:2008gf},
which are predicted in the classic secondary infall models \cite{1984ApJ...281....1F,1985ApJS...58...39B}. However, in real space significant density enhancements 
(infall caustics) do not occur, since large random motions among the clumpy, infalling material prevent their formation \cite{Diemand:2008gf}.
Cold infall is expected to occur only in the first halos forming at the bottom of the CDM hierarchy. The resulting caustics might propagate
into larger halos at lower redshifts \cite{Afshordi:2008mx}.

\section{Effects of baryons}\label{baryons}

There is a lot of work remaining to be done to quantify the effects that baryons can have in modifying the distribution of dark matter. Simulations become more complex and expensive and we do not as yet have a clear understanding of how galaxies form and evolve.

\subsection{Adiabatically steepening the density profile}

The dark matter density profiles can steepen through the adiabatic contraction due to dissipating baryons 
\cite{1980ApJ...242.1232Y,Blumenthal:1985qy,Gnedin:2004cx,Sellwood:2005pq}. The strength of this effect depends on the baryonic fraction that slowly dissipates via radiative cooling. However, accretion of baryons via cold flows may dominate the growth of many galaxies \cite{Keres:2004cq}, thus it is not yet clear how strongly this changes the inner distribution of dark matter in galaxies. For a halo that cools the cosmologically available baryons into a disk component, the dark matter density at a few percent of the virial radius increases by about a factor of two and the final, total (baryons and dark matter) density profile can resemble an isothermal sphere - comparable to observed constraints on elliptical galaxies \cite{Gnedin:2004cx}. For isolated galaxies less massive than the Milky Way, the baryon fraction decreases rapidly, $M_{baryon}\propto V_{vir}^4$, such that the smallest galaxies have captured and cooled less than 10\% of the available baryons \cite{Mayer:2003fs}. 

\subsection{Dynamically flattening the central cusp}

At smaller radii, the growth of supermassive black holes or central nuclei can steepen or shallow the very central dark matter cusp
depending of whether these structures grow adiabatically or through mergers. 
Gondolo and Silk (1999) \cite{Gondolo:1999ef} explored the effects of slow central black hole formation on the CDM cusp in the context of an enhanced annihilation signal. This mechanism can create isothermal cusps ($\rho \propto r^{-2}$) on parsec scales, with a boost factor of several orders of magnitude. If a dense star cluster surrounds the black hole (as in the center of the Milky Way) a shallower cusp ($\rho \propto r^{-1.5}$) is expected \cite{2004PhRvL..92t1304M,2004PhRvL..93f1302G}. On the other hand, if supermassive black holes form via mergers, then dark matter particles (e.g. WIMPs) can be ejected from the central halo region via three body encounters. This process could produce a constant density region of dark matter with a mass deficit roughly proportional to the mass of the sinking binary objects \cite{2002PhRvL..88s1301M}. Similar behavior would result from the formation of central stellar nuclei in galaxies. Dissipative growth would increase the central dark matter density, but formation via merging of existing star clusters would lead to an inner core \cite{Goerdt:2008pw}. A similar mechanism was studied in the context of cluster halos, whereby energy transfer to the dark matter background from dynamical friction acting on massive satellite galaxies, gave rise to an inner region with constant dark matter density \cite{ElZant:2003rp}. All of these processes have yet to be studied in a realistic cosmological context.

\subsection{Feedback and stirring - flattening the cusp}

Feedback from the star formation process has frequently been invoked to flatten cusps, especially in dwarf galaxies which have challenged the CDM paradigm through observations of rotation curves, stellar velocities and star-cluster kinematics. A single violent event, which somewhat unrealistically ejects a cosmological baryon fraction from the inner region, can redistribute the dark matter through a central revirialisation. However the most careful study of this process shows the effect to be modest, with a reduction in the central halo density by at most a factor of two to six \cite{Gnedin:2001ec}. More realistic SPH simulations in a cosmological context show that supernovae driven turbulent gas motions can impart sufficient energy to the dark matter to create a core as large as 400 parsecs in a Fornax sized galaxy \cite{2008Sci...319..174M}. This effect requires both a significant early central baryon fraction and for the Jean's mass to be accurately followed since bulk motions are driven by starbursts in giant molecular clouds. It will be interesting to compare these experiments with higher resolution adaptive mesh techniques including the effects of reionisation.

Over half of disk galaxies have stellar bars which can transfer angular momentum to dark mater particles through orbital resonances and dynamical friction. The magnitude of this process has been debated in the literature \cite{Weinberg:2005gp,2008ApJ...679..379S,Klypin:2008ay}. However even when a rigid perturber mimicking a bar was placed at the center of a CDM halo, it only affected the dark matter particles within $\sim 0.001 r_{\rm vir} \sim 300$ pc in our Galaxy. The most recent highest resolution study of this process demonstrates that the effects of bars on the central dark matter distribution is negligible \cite{Dubinski:2008yi}.

\subsection{Halo shape}

The shapes of dark matter halos can be modified by galaxy formation\cite{1991ApJ...377..365K,Kazantzidis:2004vu}: Box orbits supporting the triaxial configurations become deformed by the growing disk \cite{Debattista:2007yz}. Since particles move on eccentric orbits with a typical apocentric to pericentric distance of 6:1, halos can be visibly affected out to half the virial radius, and become almost spherical close to the galaxy. The change in shape depends on the central baryonic fraction, which is highest for elliptical galaxies. It is quite low for galaxy clusters and dwarf galaxies, their halos should therefore barely be affected. 
The detailed modification of particle orbits within the disk region has yet to be explored but could also change the
predictions for direct detection experiments \cite{Debattista:2007yz}.

\subsection{Substructure}

Galaxy formation also leads to the accretion of gas, stars and dark matter from satellites
into the disk: systems on roughly co-planar orbits suffer dynamical friction against the disk, which brings them into
the disk plane where they are disrupted \cite{Meza:2004ig,Read:2008fh}. This process produces a {\em dark matter disk}, which could
contribute a significant fraction of the local dark matter density \cite{Read:2008fh}.
Satellites on more polar orbits on the other hand will pass through the disk and experience a quick increase of tidal forces during the passage,
a so called tidal shock. Such shocks can lead to substantial mass loss, their impact depends strongly on the orbit of a subhalo and on 
its mass profile \cite{2001ApJ...559..716T,Kuhlen:2007ku}.

Small, nearby substructures will also lose significant amounts of mass as they suffer heating from individual stars \cite{Zhao:2005mb,Goerdt:2006hp}. For the smallest substructures with sizes smaller than a few hundred parsecs, impulsive collisional heating due to encounters with disk stars dominates their mass loss. Over a Hubble time most of their particles will be lost into small streams, although often an inner dense and strongly bound core is able to survive \cite{Goerdt:2006hp}.

The importance of many of these processes remain to be quantified which will be an area of activity whilst simulators attempt to create realistic galaxies from cosmological initial conditions. They will also play an important role in precision cosmology in which future observational missions plan to measure the evolution of the power spectrum, or the mass distribution on large scales to percent level precision, which requires a detailed knowledge of how baryons affect the global properties of their halos.

\section{Predictions for direct \& indirect detection experiments}

In this section we will briefly summarize the main predictions from cosmological N-body simulations for direct and indirect
dark matter detection experiments. Some of the popular dark matter candidates (e.g. some of the WIMPs)
would be directly detectable thanks to their tiny but just significant enough probability to interact with ordinary matter
(e.g. atomic nuclei in an underground detector). Indirect detection experiments look for 
$\gamma$-rays, charged particles or neutrinos produced as two WIMPs annihilate each other \cite{1996PhR...267..195J}.

\subsection{$\gamma$-rays from dark matter annihilation}\label{sec:indirectdetection}

For the NFW density profile and its variants discussed above, the total luminosity from dark matter annihilation in a halo scales like
\begin{equation}\label{lscaling}
L \propto \int \rho^2 dV \propto c_V^2 \rvmax^3 \propto \vmax^4/\rvmax \propto \vmax^3 \sqrt{c_V} \; .
\end{equation}
The half light radius is $r_{1/2} \simeq$ 0.07 $\rvmax$ for the best fit profiles in Figure \ref{denpros},
and 0.09 $\rvmax$ for the shallower NFW profile.
Combining Eqn. \ref{lscaling} with the steep velocity functions found for field halos \cite{Reed:2004vg}
and subhalos \cite{Ghigna:1998vn,Reed:2004vg,Diemand:2006ik,Diemand:2008in} $N(>\vmax) \propto d N/ d {\rm log} \vmax \propto \vmax^{-3}$,
implies that small field halos and subhalos emit more $\gamma$-rays per decade in halo size
($\vmax$ or mass) than larger ones. In the latest galaxy halo simulations all resolved subhalos
together are about as luminous as the main halo \cite{Diemand:2006ik,Diemand:2008in,Springel:2008cc}, i.e. the 
boost factor $B = L_{\rm total} / L_{\rm main halo}$ is at least two for CDM galaxy halos.
Extrapolations down to the smallest CDM subhalos increase the boost to $B = 4$ to $16$ [\cite{Diemand:2008in}],
where we assume small mass cut-offs between $10^{-12}$ and $1.0 \msun$ and we
take the nearly constant halo
concentrations on the smallest scales into account (see \ref{concentrations}).
These boost factors imply that small scale structure ($< 10^6 \msun$) dominates the diffuse extragalactic
dark matter annihilation signal 
(and also the diffuse galactic signal, except towards the Galactic center \cite{2008MNRAS.384.1627P,Kuhlen:2008aw}).
The flux from small scale clumps is proportional to 
the dark matter density averaged over larger scales (Figure \ref{radial}).

The spatial and spectral distribution of diffuse gamma rays was measured by EGRET \cite{Hunger:1997we} and
it is consistent with a superposition of (poorly constrained) astrophysical diffuse components\cite{Strong:2004de}.
Even under the unrealistic assumption of a perfect subtraction of astrophysical diffuse
foregrounds, the DM detection window of the recently launched
Fermi satellite (formerly know as GLAST) in the diffuse galactic component is quite small \cite{Baltz:2008wd}.
However, the significantly improved spatial resolution of Fermi relative to EGRET might allow it to
detect gamma rays from dark matter annihilation in subhalos, both in dwarf galaxies and in smaller,
dark subhalos \cite{2008MNRAS.384.1627P,Kuhlen:2008aw,Baltz:2008wd}. Small scale clumpiness within subhalos
increases their signal slightly, which makes a somewhat larger number of subhalos detectable \cite{Kuhlen:2008aw}. 

\subsection{Nearby dark matter distribution and charged particles from dark matter annihilation}

Besides $\gamma$-rays, dark matter annihilation would produce charged
particles and anti-particles that, due to to magnetic field entanglement,
propagate over much smaller distances within the Galaxy.
Using the local subhalo abundance from Figure \ref{radial} and extrapolating down to micro-subhalo scales
one finds that nearby subhalos produce a total flux of 40\% of the local smooth halo signal\cite{Diemand:2008in}.
In other words the local boost factor is 1.4, the uncertainty from the extrapolation is about $\pm 0.2$.
Explaining the positron excess measured by HEAT \cite{Beatty:2004cy} 
and PAMELA \cite{Adriani:2008zr} with
local dark matter annihilation requires significantly larger ($\simeq 10^4$)
enhancements \cite{Lavalle:1900wn,Bergstrom:2008gr}.
When a relatively large subhalo happens to lie within 1 kpc, the local boost factor increases,
but much larger values are unlikely: Only 5.2 percent of all random realizations have a boost factor of 3 or larger (caused by a
$\vmax \ge 3.4 \kms$ clump within 1 kpc). In only 1.0 percent of the
cases the boost factor reaches 10 or higher due to a nearby, large $\vmax \ge 5.6 \kms$ subhalo \cite{Diemand:2008in}.

\subsection{Local dark matter distribution, direct detection and capture in the sun}\label{direct_detection}

Most of the local dark matter is in a smooth component \cite{Diemand:2008in,Springel:2008cc}, the probability
that the solar system currently passes through a subhalo are quite small, even when the smallest micro-subhalos
are taken into account \cite{Diemand:2005vz,Kamionkowski:2008vw}. The large number of overlapping streams
in the inner halo also leads to rather smooth local velocity distribution functions
\cite{Moore:2001vq,Helmi:2002ss,Zemp2008}. Even the most prominent steams 
apparent in Figure \ref{psd} account for less then one percent of the local dark matter density, i.e. even if we
happened to be located within such a stream today, the bulk of detected dark matter particles would still come from the "hot" background.
Whilst current cosmological simulations are able to probe the local density and velocity distributions on kpc scales \cite{Zemp2008},
alternative methods are required to study finer structures \cite{Stiff:2003tx,Fantin:2008ur}. Further studies are needed to
quantify or exclude the relevance for dark matter detection experiments
of possible very fine grained features in the local six dimensional dark matter distribution.

At 8 kpc from the center of galaxy scale {\em pure} CDM halos the velocity distributions are peaked (positive kurtosis), because of the
shallower than isothermal potential of pure CDM halos (see \ref{shapes}). The resulting
excess of slow and fast particles relative to the Gaussian standard halo model, however, is too small to change the
interpretation of direct detection results significantly \cite{Fairbairn:2008gz}.
Galaxy formation likely changed the local velocity distribution significantly (c.f. section \ref{baryons}):
For example, the constant Milky Way rotation
curve implies an isothermal potential, which suggest a more Gaussian shape for the real velocity distribution
of local dark matter particles.

The shape of the local velocity ellipsoid correlates with the shape of the halo: it is radially anisotropic on the major axis and
tangentially on the minor axis \cite{Zemp2008}. However, while pure CDM halos are
elongated \cite{Allgood:2005eu,Kuhlen:2007ku}, the shape of the local Milky Way halo is expected to be fairly
round \cite{1991ApJ...377..365K,Kazantzidis:2004vu,Debattista:2007yz}. The {\em dark matter disk}  (c.f. Section \ref{baryons}) is
probably the most drastic deviation from the standard halo model, and it may well have
significant implications for dark matter detection \cite{Read:2008fh,Bruch:2008rx}.

The time averaged dark matter capture rate relevant for neutrino production in the center of the Sun (and Earth)\cite{1996PhR...267..195J} smears
out the (small) variations due to local clumps and streams. In pure CDM simulations the rate
is close to the rate obtained from the standard halo, just slightly higher due to the small excess of low velocity
particles relative to a Gaussian \cite{Fairbairn:2008gz}. As for direct detection, the presence of accreted
dark matter in the Milky Way's disk \cite{Read:2008fh,Bruch:2008rx} is likely the most relevant deviation
from the standard halo model and further studies are required to better understand
the properties of dark disks.

\section{Summary and Future Prospects}
\label{sec:conclusions}

Numerical simulations provide a reliable method of calculating
the evolution of the dark matter distribution and they make robust
predictions for the clustering of the dark matter component.
More recently, but less reliably, simulations have begun to resolve the
combined growth of dark matter and baryonic fluctuations.
However there is still a lot to learn; the detailed interaction between
the dark matter
and baryons is poorly understood and many comparisons with observations
and predictions
for experimental searches rely on extrapolations well below the current
resolution scale.

Whilst discrepancies between observations and simulations have been the
subject of much debate in the literature, the details of the galaxy
formation process needs to be resolved in order to fully test the CDM
paradigm. Whatever the true nature of the dark matter particle is, it must
not be too different from a cold neutralino like particle to maintain all
the successes of the model in matching large scale structure data and the
global properties of halos which are mostly in good agreement with
observations.

Even as simulations have gone past the "billion particle halo" goal, there
are still several reasons why it is interesting to further increase this
resolution. For example,
to-date the best resolved CDM satellite capable of hosting a dwarf galaxy
contains just a few million particles and its mass distribution is
unresolved on scales below a few hundred parsecs. This is the scale within
which the observations will constrain the dark matter distribution through
proper motions. Making predictions for the phase space structure and
substructure within the dark matter distribution at the position of the
sun within the Galaxy is important for direct and indirect detection
experiments. Resolving the density profile of a gravitational collapse as
the radius goes to zero is a fascinating dynamical problem that will guide
theoretical models towards understanding what processes shape the inner
structure of dark matter halos.

Ongoing surveys such as SEGUE \cite{2009arXiv0902.1781T} and RAVE \cite{2006AJ....132.1645S}
and future surveys such as GAIA and SIM-Lite
will provide us with detailed measurements of the dynamics and assembly
history of the Galaxy, its stellar halo and satellite galaxies. The 3-d
kinematics of stars from precision proper motions will constrain the
shape, orientation, density law and lumpiness of the Galactic halo. They
will measure the orbits of the Galactic satellites and the angular
momentum and orbital anisotropy of the halo stars and globular clusters to
the outer reaches of the halo. The motions of stars in the cores of nearby
dSph satellite galaxies will constrain their phase space density and dark
matter distributions.
Realistic simulations of the formation of the Galaxy and its stellar
components are urgently needed to guide these missions and to aid in their
interpretation.

We often hear that this is the precision era of cosmology. Indeed, future
precise observations of
the mass function of dark matter halos \cite{Tinker:2008ff},
high redshift 21cm tomography \cite{Mao:2008ug}, galaxy surveys such as Pan-STARRS,
LSST or the Dark Energy Survey and many more, all aim to determine the
cosmological parameters to a level that probes fundamental physics.
However there are several problems which arise when numerical simulations
are used to infer weakly or highly non-linear processes: Different
simulation codes and initial conditions software lead to differences in
results at the ten percent level \cite{Heitmann:2004gz}.
Improving the predictability of simulations by a factor of ten will require
a significant investment of collaborative work. We also need to supply
large numbers of large-volume cosmological simulations with a high dynamic
resolution that have numerical and resolution effects fully under control.
Finally, we are left with large uncertainties in the role that baryons
play in changing some of the basic properties of dark matter halos,
effects that need to be understood before we can proceed in using
simulations to probe fundamental physics rather than testing our
understanding and modeling of gas-dynamical processes.

In the past decades and for the foreseeable future, simulations
of galaxy formation do require guidance from observations. In the
coming years we will have ground and space based telescopes like ALMA or
the JWST which will reveal details of the galaxy assembly process
occurring at high redshifts. Indeed, one of the ultimate goals of
simulations is to recreate a single galaxy, or more optimistically, an
entire cluster or large volume of well resolved galaxies. The observations
are already far ahead of the predictive power of numerical simulations,
which have yet to be able to model the formation of a single pure disk
galaxy within a cold dark matter halo. Most likely this is a problem with
resolution, algorithms and sub-grid physical processes, which are always
present in such simulations. More speculatively, this failure could
reflect a problem with our standard cosmological paradigm.

\acknowledgments 
JD acknowledges support from NASA through a Hubble Fellowship grant HST-HF-01194.01,
from KITP  through National Science Foundation Grant No. PHY05-51164.

\vfil

\newpage
\bibliography{minirev}

\end{document}